\documentclass[sigconf]{acmart}

\usepackage{booktabs} % For formal tables
\usepackage{etex}
\usepackage{enumitem}
\usepackage{hyperref}
\usepackage{pgfplots}
\usepackage{xspace}
\usepackage[T1]{fontenc}
\usepackage{listings}
\usepackage{xcolor}
\usepackage{tikz}
\usepackage{color}
\usepackage{multirow}
\usepackage{hhline}
\usepackage{multicol}
\usepackage{colortbl}
\usepackage [autostyle, english = american]{csquotes}
\MakeOuterQuote{"}
\definecolor{darkgreen}{HTML}{3C8031}

\newcommand{\checkNum}[1]{{#1}}

\newcommand\tool[1]{{\textnormal{\textsc{#1}}}\xspace}

\newcommand{\csC}[0]{{{Comparison scenario}}\xspace}
\newcommand{\cs}[0]{{{comparison scenario}}\xspace}
\newcommand{\ao}[0]{\textit{AO}\xspace}
\newcommand{\an}[0]{\textit{AN}\xspace}
\newcommand{\cm}[0]{\textit{LinOS}\xspace}

\newcommand{\lineage}[0]{\tool{LineageOS}}
\newcommand{\catGreen}[0]{Possible\xspace}
\newcommand{\catRed}[0]{Not Possible\xspace}
\newcommand{\catYellow}[0]{Might Be Possible\xspace}

\newcommand{\vcell}[1]{\rotatebox[origin=c]{90}{#1}}

\definecolor{mygreen}{HTML}{48D373}
\definecolor{myred}{HTML}{EA0D30}
\definecolor{myyellow}{HTML}{FCC168}
\newcommand{\gcell}[0]{\cellcolor{mygreen}\catGreen}
\newcommand{\rcell}[0]{\cellcolor{myred}\catRed}
\newcommand{\ycell}[0]{\cellcolor{myyellow}\catYellow}

\definecolor{dkgreen}{rgb}{0,0.6,0}
\definecolor{gray}{rgb}{0.5,0.5,0.5}
\definecolor{mauve}{rgb}{0.58,0,0.82}

\usepackage[absolute,overlay]{textpos}
\usepackage{tikz}
\usepackage{xcolor}
\usepackage{calc}

\newcounter{findingctr}

\newcommand{\finding}[2]{\refstepcounter{findingctr}\emph{Finding \thefindingctr:\label{finding:#1}}}

\newlength{\boxw}
\newlength{\boxh}
\newlength{\shadowsize}
\newlength{\boxroundness}
\newlength{\tmpa}
\newsavebox{\shadowblockbox}

\newenvironment{findingenv}[1]%
{\vspace{0.2cm}\noindent
\begin{lrbox}{
\shadowblockbox
}
\begin{minipage}{.98\columnwidth}
\finding{#1}~}%
{\end{minipage}\end{lrbox}%
\settowidth{\boxw}{\usebox{\shadowblockbox}}%
\settodepth{\tmpa}{\usebox{\shadowblockbox}}%
\settoheight{\boxh}{\usebox{\shadowblockbox}}%
\addtolength{\boxh}{\tmpa}%
\begin{tikzpicture}
\addtolength{\boxw}{\boxroundness * 2}
\addtolength{\boxh}{\boxroundness * 2}

\foreach \x in {0,.05,...,1}
{
\setlength{\tmpa}{\shadowsize * \real{\x}}
\fill[xshift=\shadowsize - 1pt,yshift=-\shadowsize + 
1pt,black,opacity=.04,rounded corners=\boxroundness] 
(\tmpa, \tmpa) rectangle +(\boxw - \tmpa - \tmpa, \boxh - \tmpa - 
\tmpa);
}

\filldraw[fill=white!50, draw=black!80, rounded corners=\boxroundness] (0, 
0) rectangle (\boxw, \boxh);
\draw node[xshift=\boxroundness,yshift=\boxroundness,inner sep=0pt,outer 
sep=0pt,anchor=south west] (0,0) {\usebox{\shadowblockbox}};
\end{tikzpicture}\vspace{0cm}%
}
\makeatletter
\renewcommand\paragraph{\@startsection{paragraph}{4}{\parindent}%
  {-.1\baselineskip \@plus -2\p@ \@minus -.2\p@}%
  {-1.5\p@}%
  {\@parfont\@adddotafter}}
\makeatother

\begin{document}
\title{The Android Update Problem: An Empirical Study}

\author{Mehran Mahmoudi and Sarah Nadi}
\affiliation{%
  \institution{University of Alberta, Edmonton, Alberta}
}
\email{{mehran,nadi}@ualberta.ca}

\begin{abstract}
Many phone vendors use Android as their underlying OS, but often extend it to add new functionality and to make it compatible with their specific phones.
When a new version of Android is released, phone vendors need to merge or re-apply their customizations and changes to the new release. 
This is a difficult and time-consuming process, which often leads to late adoption of new versions.
%Ideally, automated support that can merge the vendor-specific changes with the changes that happened in the new release would speed up the process.
In this paper, we perform an empirical study to understand the nature of changes that phone vendors make, versus changes made in the original development of Android. 
By investigating the overlap of different changes, we also determine the possibility of having automated support for merging them.
We develop a publicly available tool chain, based on a combination of existing tools, to study such changes and their overlap.
As a proxy case study, we analyze the changes in the popular community-based variant of Android, \lineage, and its corresponding Android versions.
We investigate and report the common types of changes that occur in practice. 
%~\sn{one sentence about interesting changes?}
Our findings show that \checkNum{83\%} of subsystems modified by \lineage are also modified in the next release of Android.
By taking the nature of overlapping changes into account, we assess the feasibility of having automated tool support to help phone vendors with the Android update problem.
Our results show that \checkNum{56\%} of the changes in \lineage have the potential to be safely automated. %We discuss concrete tooling suggestions for this automation.

%Many phone vendors use Android as their underlying OS, but often extend it to add new functionality and to make it compatible with their specific phones. When a new version of Android is released, phone vendors need to merge or re-apply their customizations and changes to the new release. This is a difficult and time-consuming process, which often leads to late adoption of new versions. In this paper, we perform an empirical study to understand the nature of changes that phone vendors often make, versus changes made in the original development of Android. Using this insight, we also determine the feasibility of having automated support for merging these changes. We study multiple versions of two community-based customized variants of Android, namely CyanogenMod and Paranoid Android, and their corresponding Android versions. We discuss the nature of the changes we discovered and their overlaps, and we assess them to identify potentially conflicting changes. Our results show that 53% of the changes in CyanogenMod and 47% of the changes in Paranoid Android have the potential to be safely automated.
\end{abstract}

\copyrightyear{2018} 
\acmYear{2018} 
\setcopyright{acmcopyright}
\acmConference[MSR '18]{MSR '18: 15th International Conference on Mining Software Repositories }{May 28--29, 2018}{Gothenburg, Sweden}
\acmBooktitle{MSR '18: MSR '18: 15th International Conference on Mining Software Repositories , May 28--29, 2018, Gothenburg, Sweden}
\acmPrice{15.00}
\acmDOI{10.1145/3196398.3196434}
\acmISBN{978-1-4503-5716-6/18/05}




% \settopmatter{printacmref=false}
% \renewcommand\footnotetextcopyrightpermission[1]{} % removes footnote with conference information in first column

\maketitle

\section{Introduction}
% Android
Google's open-source mobile operating system (OS), Android, is used by the majority of phone vendors \cite{gartner} and currently has approximately 80\% of the market share of smart phones around the world~\cite{mobileStats}.
Using the Android Open Source Project (AOSP), phone vendors are able to access Android's source code, and can implement their own device specifications and drivers \cite{AndroidArchitecture}. 
Since Android is open source, phone vendors can add their own enhancements, including new hardware capabilities and new software features.

%Problems with modification
When a new version of AOSP is released, phone vendors need to obtain the new version and re-apply their modifications to it. Due to the complexity of this task, the majority of devices that use Android may not run on the most recent version right away. Based on data collected by Google in July 2017~\cite{AndroidVersions}, \checkNum{27\%} of Android-based devices run an Android version that is at least three years old, which is especially problematic for security updates~\cite{thomas:security, thomas:lifetime}.

%What do we want to do
The process of re-applying changes from an independently modified version of Android to a newer version of Android is a time-consuming and manually intensive task. 
This process can be viewed as a general software merging problem: we want to merge the changes made by vendors with the changes made by Android developers. However, unlike common merge scenarios in practice, there is no systematic reuse mechanism like branching or forking: vendor modifications are applied in a different repository independent of changes made by Android developers. 
%While this can be viewed as a software merging problem, these modifications typically happen over long periods of time.
%This may mean that the vendor-specific version and the AOSP version may have significantly diverged such that automated merging is no longer feasible.
%Alternatively, the same parts of the code could have been heavily modified on both sides that automated merging would simply lead to lots of conflicts, rendering it useless.
%Camera Ready Commented
%This process is also related to Application Programming Interface (API) migration. In essence, the vendor-specific version of Android uses the AOSP APIs so changing them should be a matter of updating the client calls. 
%While lots of research effort has been dedicated to the API evolution and migration problem (e.g.,~\cite{Xing:2005:UAO:1101908.1101919,Kim:2007:AIS:1248820.1248866,Dagenais:2008:RAC:1368088.1368154}), it is not entirely applicable in this case since the vendor-specific version is not simply a client of the AOSP API, but rather a modified version of it.
While developers can use a mix of Application Programming Interface (API) migration and software merging tools to help them with the process, we are not aware of any single off-the-shelf tool that can be used to automatically accomplish this merging task. 
However, before attempting to automate this task, we argue that we first need to understand the changes that vendors make versus those that Android developers make, and whether these changes overlap.
%In this paper, we take the viewpoint that consolidating vendor-specific changes with the new Android changes is a software merging task. We study the evolution of vendor-specific code and Android code to understand the feasibility of automating this merging task. However, instead of simply trying to merge the changes and then analyze textual conflicts, 

In this paper, we focus on understanding the \textit{nature} of the changes that occur in a large software system such as Android when compared to an independently modified version of it. 
If we understand the nature of the changes on a semantic level (e.g., add method argument), we can identify current tools and techniques that can address them, or identify technology gaps that need to be filled. We focus on the Java parts of AOSP since Google recently announced the introduction of Project Treble~\cite{GoogleAnnounceTreble,ProjectTreble} which allows easier update of the hardware-specific parts implemented in C through a new architecture of these layers. However, this does not solve the problem of vendor-specific Java changes in AOSP itself.

%Camera Ready Commented
%The Android update problem is a prevalent problem for phone vendors which led Google itself to look for a solution.
%Google recently announced the introduction of Project Treble~\cite{GoogleAnnounceTreble,ProjectTreble}, which would separate the vendor-specific, hardware parts of Android, typically implemented in C, from the rest of the AOSP, typically implemented in Java.
%This allows easier update of the hardware-specific parts, or more precisely, it removes the need to re-apply any AOSP changes to them.
%However, this does not solve the problem of vendor-specific Java changes in AOSP itself.
%Given that Google is already coming up with a solution for the C-based vendor-specific changes, in this paper, we focus on the Java parts of AOSP.

Since we do not currently have access to proprietary vendor-specific code, we use a popular community-based variant of Android, \lineage, as a proxy for a vendor-based version of Android. 
For each subsystem in AOSP, we track the method-level changes in the source code using a combination of existing code-evolution analysis tools: SourcererCC~\cite{Sajnani:2016:SSC:2884781.2884877}, ChangeDistiller~\cite{changedistiller}, and RefactoringMiner~\cite{RefactoringMiner}. 
We are interested in changes in two directions.
First, between an old version of Android and its subsequent new version.
Second, between the old version of Android and the independently modified \lineage version that is based on that old Android version.
We use the term \textit{\cs} to describe the combination of these three versions.
For each \cs, we first analyze the types of changes that have occurred.
We then analyze the intersection of the two computed sets of changes to estimate the proportion of changes that can potentially be automated.  
Specifically we answer the following research questions:

% \vspace{-0.05cm}
%summary of results
\begin{enumerate}[label=\textbf{RQ\arabic*}]
\item \textit{Which parts of Android are frequently modified in AOSP vs. \lineage?}
In each \cs analyzed, there are at least \checkNum{two} common subsystems between \lineage and Android in their corresponding lists of top \checkNum{five} most-changed subsystems. 
This suggests that both systems often need to apply many changes to the same subsystems. 
%Camera Ready Commented
%\textit{SystemUI} and \textit{Settings} are among the top \checkNum{five} most-changed subsystems in both AOSP and \lineage, for \checkNum{seven} out of \checkNum{eight} \cs{s}.

\item \textit{What are the overlapping types of changes between AOSP and \lineage modifications?}
We find that the majority of \lineage changes (avg: \checkNum{50.19\%}, median \checkNum{49.46\%}) change a given method body, while Android does not modify that method. 
On the other hand, we find that an average of \checkNum{16.08\%} (median: \checkNum{16.20\%}) of \lineage changes modify a given method body that Android also modifies in the new version. 

\item \textit{How feasible is it to automatically re-apply \lineage changes to AOSP?}
By considering the semantics of the different changes, we find that on average, \checkNum{56\%} of \lineage changes have the potential to be automated when integrating AOSP's changes.
However, for \checkNum{28\%} of the cases, developer input would be needed.
The automation feasibility of the remaining \checkNum{16\%} depends on the specifics of the change. %In general, this is good news for vendors since the majority of changes are non problematic, while some of the remaining ones can be addressed by current tools.

\end{enumerate}

% \vspace{-0.1cm}
In summary, the contributions of this paper are as follows:
% \vspace{-0.1cm}
\begin{itemize}
\item To the best of our knowledge, this is the first research paper to discuss the Android update problem, i.e., the problem of integrating vendor-specific changes with AOSP updates.
\item An empirical study of the changes that occur in the \checkNum{eight} latest release of \lineage versus their AOSP counterparts.    
\item A breakdown of the overlap between these changes and the feasibility of automation for each category, including a discussion of the suitability of current tooling for the task.
\item Open-source tool chain~\cite{tooling} that integrates existing tools to identify method-level change sets and automates the overlap analysis. While this paper focuses on one case study, our tool chain can be applied to other community based or vendor-specific Android variants, as well as other Java systems. %Our tooling is available online.

\end{itemize}

%The remaining of paper
%\sn{If we are tight in space, we can remove this paragraph}The rest of this paper is organized as follows: In Section \ref{background}, we explain background information required for and the motivation behind our study. Next, in Section \ref{related}, we discuss the related literature regarding the Android API and API evolution in general. In Section \ref{methodology}, we explain how we measured the possibility for automation in Android and an independently modified variant of it. In Section \ref{eval} we discuss our evaluation process and the dataset used. Section \ref{results} is dedicated to analysing the results we got from running out tool on the dataset. In Section \ref{threats} we go over the threats to our study's validity. Finally, in Section \ref{conclusion}, we conclude our work and discuss its impact and future directions.

% d
%\begin{figure}[t]
%\centering
%\includegraphics[width=0.5\textwidth]{figures/PieChart.pdf}
%\caption{Proportion of devices running different versions of Android as of July 7, 2017~\cite{AndroidVersions}}
%\label{figure:android_versions}
%\end{figure}

\section{Background} \label{background}

% \begin{table}[t!]
% \centering
% \bgroup
% \def\arraystretch{1.2}
% \caption{Corresponding Android versions in \lineage.}
% \label{table:cm_version}
% \resizebox{0.4\textwidth}{!}{
% \begin{tabular}{|c|c|c|c|} 
% \hline
% \tool{LineageOS} version & Android version & Android code name\\ \hline
% % 10.0 & 4.0 & Ice Cream Sandwich\\ \hline
% 10.1 & 4.2.2 & Jelly Bean\\ \hline
% 10.2 & 4.3.1 & Jelly Bean\\ \hline
% 11.0 & 4.4.4 & KitKat\\ \hline
% 12.0 & 5.0.2 & Lollipop\\ \hline
% 12.1 & 5.1.1 & Lollipop\\ \hline
% 13.0 & 6.0.1 & Marshmallow\\ \hline
% 14.0 & 7.0.0 & Nougat\\ \hline
% 14.1 & 7.1.2 & Nougat\\ \hline
% \end{tabular}
% }
% \egroup
% \end{table}

\paragraph{Android Architecture}
The Android source-tree architecture consists of several layers \cite{AndroidArchitecture}. Each layer has a different functionality and is written in a specific language. The lowest layer is the Linux Kernel level. Phone vendors are expected to include device drivers for their hardware in this layer. On top of the Linux Kernel, there is the Hardware Abstraction Layer (HAL). This layer defines a standard interface for hardware vendors to implement and allows Android to be agnostic to lower-level driver implementations. Phone vendors are responsible for interaction between the HAL implementation and their device drivers in the Kernel layer.

While development in HAL and the Linux Kernel are done in C/C++, higher-level layers such as the Android System Services and the Application Framework use Java. Phone vendors may apply changes to these layers for various reasons, such as adding new functionality or building their ecosystem's look and feel.

The Android source code is maintained in multiple repositories that collectively build up the source-tree. A list of these repositories is maintained in a repository called Android Platform Manifest \cite{AndroidPlatformManifest}.

\paragraph{Android Subsystems} To understand which OS parts are modified, it would be useful to divide the Android source code into different subsystems.
%Initially, we thought of considering each repository listed in the platform manifest file as an Android subsystem. However, we noticed that some repositories are significantly larger than others and include multiple sub-folders such where each of these sub-folders could be considered as an independent subsystem. Thus, simply considering each repository as a subsystem would lead to some mismatch in the intuitive notion of what a subsystem means. Because of this reason, we looked for a different criteria to define subsystems.
In application development for Android, each folder that contains an \textit{AndroidManifest.xml} file is compiled and built into a \textit{.apk} file, which is later installed on the phone. Conveniently, all parts of Android that are implemented in Java are considered as apps, meaning that they have an \textit{AndroidManifest.xml} file in their source and are packaged into \textit{.apk} files. Considering this fact, we decided to define the notion of an Android Java \textit{subsystem} as a folder that contains an \textit{AndroidManifest.xml} file.

\paragraph{\lineage} %Our final goal is to merge the modified vendor version, which is based on the old version of AOSP, with the new version of AOSP. 
Because phone vendors do not make the source code of their customized Android OS version publicly available, we need a proxy for a vendor-specific Android variant. There are a number of community-based Android OS variants available. Since most of them are open source, it is possible to use them as a proxy for a modified vendor-specific version of Android for our research purposes. \lineage~\cite{cyanogen:archived, forbes:cyanogen} is a popular alternative operating system for devices running Android. It offers features and options that are not a part of the Android OS distributed by phone vendors. These features include native Android user interface, CPU overclocking and performance enhancement, root access, more customization options over various parts of the user interface, etc.
In 2015, \tool{LineageOS} had more than 50 million active users \cite{forbes:cyanogen} and a community of more than 1,000 developers. It it actually a continuation of \tool{CyanogenMod}, a project that was discontinued in December 2016 and that continued under the new name, \lineage.

%\begin{figure}[t]
%\centering
%\includegraphics[width=7cm]{figures/android_stack}
%\caption{Android Platform Architecture~\cite{AndroidArchitecture}. Image used in accordance to Creative Commons 3.0 Attribution License.}
%\label{android_stack}
%\end{figure}

% Camera Ready Commented, we mentiond this in evaluation
%Each version of \lineage, which is a separate branch in the \lineage repository, is based on the source-tree of an AOSP version. The version each \lineage branch is based on can be found in the corresponding manifest file that contains the list of required repositories for building up the source-tree.

%Table \ref{table:cm_version} shows most recent versions of \tool{LineageOS} and their corresponding Android versions. Some versions of Android do not have a corresponding CM version, so we do not include them in the table, and do not consider them further in our work.
\section{Related Work} \label{related}
We divide related work into four categories: software evolution, API migration, software merging, and software product lines.

\textbf{Software Evolution.} A lot of work focused on identifying types of changes that occur during software evolution~\cite{SMR:SMR344}, with varying goals and underlying techniques. For example, Van Rysselberghe and Demeyer \cite{VanRysselberghe:2003:RSS:942803.943720} use clone detection to identify method moves. Xing and Stroulia~\cite{Xing:2005:UAO:1101908.1101919} develop \tool{UMLDiff}, which compares UML models to produce structural change-trees. However, they cannot detect changes within method bodies. Kim et al.~\cite{Kim:2007:AIS:1248820.1248866, Kim:2009:DRS:1555001.1555046} describe code changes between two versions of a program as first-order relation logic rules. They also cannot detect changes within method bodies.
Our work combines several existing tools and techniques, such as \tool{ChangeDistiller}~\cite{changedistiller}, \tool{SourcerCC}~\cite{Sajnani:2016:SSC:2884781.2884877}, and \tool{RefactoringMiner}~\cite{RefactoringMiner} to get a comprehensive set of method-level changes.

\textbf{API migration.} The techniques used to identify changes during software evolution have often been used as a basis for updating client projects if a library API they use changes. For example, Xing and Stroulia~\cite{Xing:2007:ASD:1314032.1314041} develop \tool{Diff-CatchUp} based on \tool{UMLDiff} to automatically recognize API changes and propose plausible replacements to obsolete APIs. Similarly, Dagenais and Robillard \cite{Dagenais:2008:RAC:1368088.1368154} develop \tool{SemDiff}, which tracks and analyzes the evolution of a framework to infer high-level changes and recommend adaptive changes to client applications. Previous work showed that API migration is a prevalent problem in Android, where app developers do not generally keep up with the pace of API updates and often use outdated APIs~\cite{McDonnell:2013:ESA:2550526.2550557}. 
Linares-V\'{a}squez et al.~\cite{Linares-Vasquez:2013:ACF:2491411.2491428} also study Android's public APIs and find out that using unstable and fault-prone APIs negatively impact the success of Android apps.
There are similar additional studies on Android public APIs~\cite{McDonnell:2013:ESA:2550526.2550557}.
Our work does not focus only on updates to public APIs used by clients, but rather involves analyzing the evolution of independently maintained versions of a piece of software.

\textbf{Software Merging.}
With branch-based or fork-based development in version-control systems, such as git, merging different branches or forks can often cause conflicts.
A lot of work has been dedicated to improving software merging to enable automatic resolution of \textit{textual} or \textit{syntactic} conflicts. We briefly discuss some recent related work, but refer the reading to  the work by Tom Mens~\cite{MensMerging} for a detailed survey on software merging.

To reduce conflicts, some researchers proposed continuously running or merging developer changes in the background to warn developers about potential conflicts before they actually occur~\cite{BrunFSE11,GuimaresMergeConflict}.
Semi-structured merge has been proposed by Apel et al.~\cite{Apel:2011SS} as a way to reduce syntactic conflicts by leveraging some of the language semantics. To identify \textit{semantic merge conflicts} that lead to build or test failure, Nguyen et al.~\cite{Nguyen:2015} propose using variability-aware execution to simultaneously run the test suite on all possible combinations of changes and identify combinations that may result in test failures. 
Our analysis lies at a level between traditional syntactic and semantic merge conflicts: we go beyond simple overlapping textual changes and look at overlapping semantic changes taking the nature of the change into account. However, we do not actually execute any merging to asses its impact on build or test failure. 

\textbf{Software Product Lines (SPLs).} The idea of consolidating independently evolved versions of a software system is also related to the SPL domain. Various studies and tools have been proposed to identify commonalities and variability between source code of multiple software versions and consolidate them in an SPL (e.g.,~\cite{Duszynski:SPL},~\cite{Fischer:ECCO},~\cite{FenskeSPLClones}). As opposed to traditional merging, the goal with SPLs is to integrate the variants but keep them configurable such that it is possible to choose any of the existing behavior.

Finally, we would like to note that there is existing work that studies techniques and requirements for providing a dynamically updatable operating system~\cite{baumann2005providing,Arnold:2009:KAR}. Obviously, if Google decides to change the Android architecture to support such dynamic updates on all levels of the OS, then the problem would be solved. However, in our work, we look at the current state of the Android OS and address the Android update problem from the perspective of phone vendors, not from the perspective of the OS owners.

\section{Methodology and Tool Chain} \label{methodology}

\begin{figure}[t!]
\centering
\includegraphics[width=0.4\textwidth]{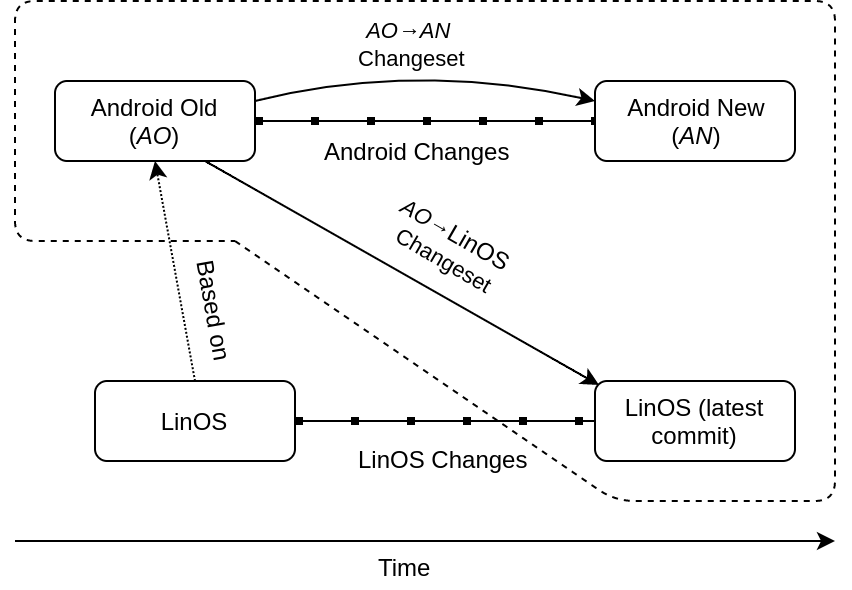}
\caption{An overview of a \cs in a given subsystem. The three versions within the dashed area are those used in the \cs.\vspace{-0.5cm}}
\label{figure:comparison_scenario}
\end{figure}

\begin{figure*}[t!]
\centering
\includegraphics[width=0.8\textwidth]{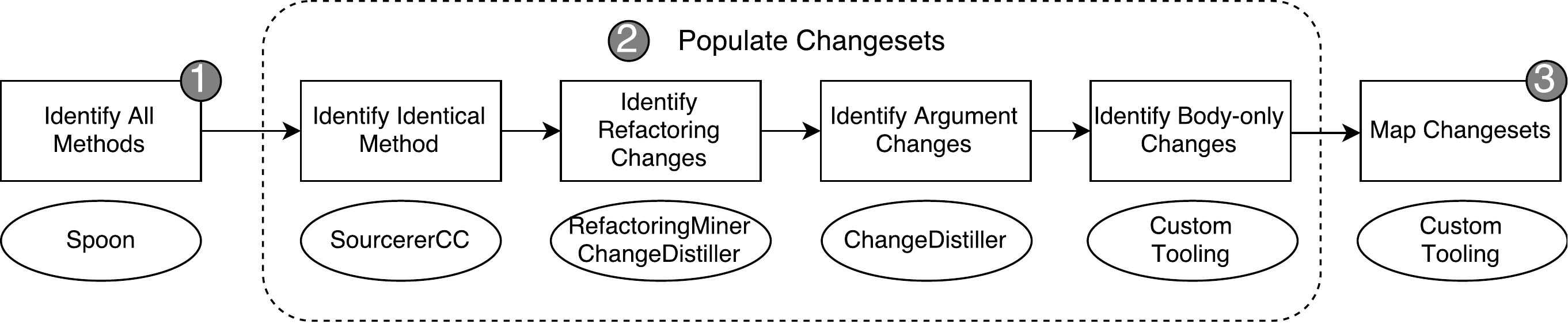}
\caption{An overview of our tool chain for identifying and comparing changes in a \cs. Each square corresponds to a step and the oval below it is the tool used in that step.\vspace{-0.5cm}}
\label{figure:methodology_overview}
\end{figure*}

Figure \ref{figure:comparison_scenario} shows the relation between the evolution of \lineage with respect to AOSP. A new \lineage version is created based on a given Android version, \textit{Android Old} (\ao) in this case. Developers then evolve this \lineage version to add their own customization to the original Android version it is based on. The final state of these customizations would be in the latest commit of \lineage (\cm). Simultaneously, Android developers are modifying Android to eventually release a new version, \textit{Android New} (\an). %\textbf{The goal of our work is to compare the changes that happened in CyanogenMod to those that happened in Android. Understanding the relationship between these changes allows later automation of applying the CyanogenMod-specific changes to AN.} 
We call the group of three versions required to compare Android changes to the \lineage changes (i.e., \ao, \an, and \cm) a \textit{\cs}. These three versions are surrounded by the dashed box in Figure~\ref{figure:comparison_scenario}. For each \cs, we identify all subsystems that belong to \ao and track their evolution through AOSP and \lineage changes.
We focus on changes at the method level, because the method-level provides a concrete and self-contained granularity level that we can reason about, and it has been used in many software evolution studies~\cite{SMR:SMR344}. 

Figure~\ref{figure:methodology_overview} shows the steps we follow in our tool chain in order to analyze a \cs. It consists of three main steps that are applied to each subsystem in the \cs.
%Recall that we consider any folder with an \textit{AndroidManifest.xml} file as a subsystem. 
We now explain each of these steps. Our tool chain, which is a combination of pertaining third-party tools and some custom tooling, is implemented in Java and Python and is available online~\cite{tooling}.

\subsection{Step 1: Identify All Methods}
We use Spoon \cite{pawlak:spoon} to generate the Abstract Syntax Trees (ASTs) of all Java files in the three versions of a subsystem. Using the ASTs, we extract a signature for each method that uniquely identifies it. %This signature includes the package name, the class name (and its outer classes if any), the return type of the method, and the method's name and type of arguments.
This gives us a set of all methods in each of the three versions of a subsystem.

\subsection{Step 2: Populate Changesets}
\label{sec:typesofchanges}

We call the set of method changes between two given versions of a subsystem a \textit{changeset}. In this step, we want to populate two changesets, one between \ao and \an and one between \ao and \cm. Our tooling receives the \ao, \an, and \cm versions of a subsystem as input and populates these two changesets. We now describe the type of changes a method can undergo and the tooling we use to detect the change.
When discussing the changes, we break down methods into signature and body. A \textit{method signature }includes the package and class it is located in, its name, arguments, and return type. The \textit{method body} is the implementation of the method. Changes to these two entities are not mutually exclusive, however, meaning that a method could be changed in both ways.

%%identical
\textbf{{Identical Methods}} are those that do not have any changes in neither their signature nor body.
To identify identical methods, we use SourcererCC~\cite{Sajnani:2016:SSC:2884781.2884877}, a token-based clone detection tool. % designed for detecting both exact and near-miss clones in large-scale projects.
It has a high reported rate of 90\% recall (100\% for certain benchmarks) and 83\% precision. We configure it with 100\% similarity threshold to look for exact clone pairs on the function level.
%For example, assume method $m1$ exists in the \ao version of a subsystem. With the above setting, SourcererCC would report any method $m2$ in \an, or \cm, that has the exact same signature and body. We then consider the relationship between $m1$ and $m2$ as an identical change and add it to the respective changeset.%, \ao to\an in this example.

%%refactoring
\textbf{{Refactoring Changes.}}
Generally speaking, a refactoring change is a change that does not semantically alter the source code.
In refactoring changes, the method signature is changed, although the body could be the same, depending on the type of refactoring. We consider the most common types of refactorings: 
\begin{itemize}
\item \textbf{Method Move:} A method's class or package is changed.
\item \textbf{Method Rename:} A method's name is changed.
\item \textbf{Method Inline:} A method is removed and its body is moved to the place it was originally called from.
\item \textbf{Method Extraction:} Part of a method ($m1$) is moved to a newly created method ($m2$), and $m2$ is called from $m1$.
\item \textbf{Argument Rename:} A method argument's name is changed.
\item \textbf{Argument Reorder:} The order of arguments is changed.
\end{itemize}

We use RefactoringMiner~\cite{RefactoringMiner} to detect refactoring changes at the method level between different versions of a subsystem. RefactoringMiner has a reported recall and precision rate of 93\% and 98\%, respectively, which is fairly high compared to other similar tools. RefactoringMiner can detect the first four refactorings in the above list, but does not detect argument rename or reorder. To detect these two types of refactorings, we use ChangeDistiller~\cite{changedistiller}, which extracts source code changes based on tree differencing. %It can detect \checkNum{48} types of changes given two versions of a source code file, but for this change type, we only use the argument rename and argument reorder changes it reports.

%%argument changes
\textbf{{Argument Changes.}}
There are other types of argument changes that would not be considered as refactoring changes. These include adding a new argument or deleting one, as well as changing an argument's type.
We use ChangeDistiller \cite{changedistiller} to detect these three types of argument changes and add them to the respective changesets under the argument change category. The reported precision and recall rates of ChangeDistiller are also fairly high, 78\% and 98\% for recall and precision respectively~\cite{changedistiller}.

\textbf{{Body-only Changes.}}
A method could have changed between two versions of a subsystem, but does not belong to any of the above categories.
We already covered changes that can only affect a method's signature (method move, method rename, argument rename, and argument reorder) or both its signature and body (method inline, method extraction, and argument changes). Body-only changes include methods that have the exact same signature between two versions of their subsystem, but have modified bodies.

For each of the two changesets, we implement our own tooling to search for methods in the \ao version of the subsystem that have not been paired with a method in the related version. Between these unmatched methods, we look for pairs with matching signatures. Such pairs of methods will definitely have different bodies; otherwise, they would be identical and would have been discovered by SourcererCC when identifying identical methods. We add such methods to the body-only change category.

\begin{table*}
\centering
\bgroup
% \vspace{-0.5cm}
\def\arraystretch{1.2}
\caption{Mappings of changesets and the feasibility of automation in a \cs}
\label{table:methods_mappings}
% \vspace{2mm}
\resizebox{\textwidth}{!}{
\begin{tabular}{cc|c|c|c|c|c|c|c|c|c|c|c|} 
\hhline{~~|-|-|-|-|-|-|-|-|-|-|-}
\multicolumn{2}{c|}{} & \multicolumn{11}{c|}{\textbf{\ao$\rightarrow$\cm} Changeset}\\ \hhline{~~|-|-|-|-|-|-|-|-|-|-|-}

%REFACTORED_MOVE	REFACTORED_RENAME	REFACTORED_INLINE	REFACTORED_EXTRACT	REFACTORED_ARGUMENTS_RENAME	REFACTORED_ARGUMENTS_REORDER	ARGUMENTS_CHANGE_ADD	ARGUMENTS_CHANGE_REMOVE	ARGUMENTS_CHANGE_TYPE_CHANGE	BODY_CHANGE_ONLY	NOT_FOUND

& & \vcell{Method Move} & \vcell{Method Rename} & \vcell{Method Inline} & \vcell{Method Extract} & \vcell{Argument Rename} & \vcell{Argument Reorder} & \vcell{Argument Add} & \vcell{Argument Remove} & \vcell{Argument Type Change} & \vcell{Body-only} & \vcell{Unmatched}\\ \hhline{|-|-|-|-|-|-|-|-|-|-|-|-|-}

\multicolumn{1}{|c|}{\parbox[t]{2mm}{\multirow{12}{*}{\rotatebox[origin=c]{90}{\textbf{\ao$\rightarrow$\an} Changeset}}}}
& Identical & \gcell & \gcell & \gcell & \gcell & \gcell & \gcell & \gcell & \gcell & \gcell & \gcell & \rcell  \\ \hhline{|~|-|-|-|-|-|-|-|-|-|-|-|-}
\multicolumn{1}{|c|}{} & Method Move & \ycell & \gcell & \rcell & \rcell & \gcell & \gcell & \gcell & \gcell & \gcell & \gcell & \rcell   \\ \hhline{|~|-|-|-|-|-|-|-|-|-|-|-|-}
\multicolumn{1}{|c|}{} & Method Rename & \gcell & \ycell & \rcell & \gcell & \gcell & \gcell & \gcell & \gcell & \gcell & \gcell & \rcell   \\ \hhline{|~|-|-|-|-|-|-|-|-|-|-|-|-}
\multicolumn{1}{|c|}{} & Method Inline & \rcell & \rcell & \ycell & \rcell & \rcell & \rcell & \rcell & \rcell & \rcell & \gcell & \rcell   \\ \hhline{|~|-|-|-|-|-|-|-|-|-|-|-|-}
\multicolumn{1}{|c|}{} & Method Extract & \rcell & \gcell & \rcell & \ycell & \gcell & \gcell & \rcell & \rcell & \rcell & \rcell & \rcell   \\ \hhline{|~|-|-|-|-|-|-|-|-|-|-|-|-}
\multicolumn{1}{|c|}{} & Argument Rename & \gcell & \gcell & \rcell & \gcell & \ycell & \gcell & \ycell & \ycell & \gcell & \gcell & \rcell   \\ \hhline{|~|-|-|-|-|-|-|-|-|-|-|-|-}
\multicolumn{1}{|c|}{} & Argument Reorder & \gcell & \gcell & \rcell & \gcell & \gcell & \ycell & \rcell & \rcell & \gcell & \gcell & \rcell   \\ \hhline{|~|-|-|-|-|-|-|-|-|-|-|-|-}
\multicolumn{1}{|c|}{} & Argument Add & \gcell & \gcell & \rcell & \rcell & \ycell & \rcell & \ycell & \rcell & \rcell & \rcell & \rcell   \\ \hhline{|~|-|-|-|-|-|-|-|-|-|-|-|-}
\multicolumn{1}{|c|}{} & Argument Remove & \gcell & \gcell & \rcell & \rcell & \ycell & \rcell & \rcell & \ycell & \rcell & \rcell & \rcell   \\ \hhline{|~|-|-|-|-|-|-|-|-|-|-|-|-}
\multicolumn{1}{|c|}{} & Argument Type Change & \gcell & \gcell & \rcell & \rcell & \gcell & \gcell & \rcell & \rcell & \ycell & \rcell & \rcell   \\ \hhline{|~|-|-|-|-|-|-|-|-|-|-|-|-}
\multicolumn{1}{|c|}{} & Body-only & \gcell & \gcell & \gcell & \rcell & \gcell & \gcell & \rcell & \rcell & \rcell & \ycell & \rcell   \\ \hhline{|~|-|-|-|-|-|-|-|-|-|-|-|-}
\multicolumn{1}{|c|}{} & Unmatched & \rcell & \rcell & \rcell & \rcell & \rcell & \rcell & \rcell & \rcell & \rcell & \rcell & \rcell   \\ \hhline{|-|-|-|-|-|-|-|-|-|-|-|-|-}

% \multicolumn{1}{|c|}{} & Argument Change & \cellcolor{myyellow}Potential Conflict & \cellcolor{myred}Conflict & \cellcolor{myyellow}Potential Conflict & \cellcolor{myred}Conflict \\ \hhline{|~|-|-|-|-|-}

% \multicolumn{1}{|c|}{} & Body-only & \cellcolor{myyellow}Potential Conflict & \cellcolor{myyellow}Potential Conflict & \cellcolor{myyellow}Potential Conflict & \cellcolor{myred}Conflict \\ \hhline{|~|-|-|-|-|-}

% \multicolumn{1}{|c|}{} & Deleted & \cellcolor{myred}Conflict & \cellcolor{myred}Conflict & \cellcolor{myred}Conflict & \cellcolor{mygreen}No Conflict \\ \hhline{|-|-|-|-|-|-}

\end{tabular}
}
\egroup
\end{table*}

\textbf{{Unmatched Methods.}}
Given two versions of a system, a method that exists in the old version might have undergone large changes that prevent existing tools from matching it with the corresponding method in the new version. The method could have also been simply deleted. We use a simplified heuristic where if we exhaust our search for a match for a method by going through all the above change types, we categorize this method as unmatched.

%\subsubsection{{New Methods.}}
%Similar to deleted methods, a method may not exist in the old version, but is added to the new version. After we finish matching methods in \an and \cm with corresponding methods in \ao, we assume that the remaining unmatched methods in \an and \cm are new methods that are added in those versions.

\subsection{Step 3: Map Changesets}
\label{sec:mapping-analysis}
After acquiring the two changesets \ao to \an and \ao to \cm, we develop custom tooling that maps the changes with similar signatures in \ao from one changeset to the other.
This mapping allows us to understand the extent and nature of overlapping changes. 
Table~\ref{table:methods_mappings} shows the potential overlapping changes to the same method, by considering all the possible combinations of changes.
Column and row labels represent types of changes in each changeset.%, as explained in Section \ref{sec:typesofchanges}.

The next step is to understand if integrating each category of overlapping change can be automated.
%For example, if a method in \ao is refactored in \cm while it is deleted in \an, we cannot automatically apply that change in \cm to its corresponding method in \an. However, if it has remained identical in \an, it is possible to automatically apply this refactoring change to the corresponding method in \an. 
To determine this, we consider the nature of the changes and categorize each cell based on the possibility of automatically integrating both changes. Green cells indicate that automation is possible. Red cells indicate cases where the developer's input is needed. Yellow cells indicate potential problems: depending on the details of the change, automation may or may not be possible.
We have not included the identical change type between \ao and \cm in the table, because there are no changes to apply on \an for those methods.
It is worth mentioning that by discussing automation possibility, we are not suggesting that an off-the-shelf tool exists that can already fully automate the integration or the merge. We are suggesting that given the nature of the changes and state-of-the-art tools, it is possible to develop a tool chain that utilizes a combination of techniques in order to automatically merge these change types.    

We start by discussing our reasoning across the first row, last row, and last column of Table~\ref{table:methods_mappings} as they have the same color across. If a method in \an is identical to \ao, then we can safely automatically apply any change from \cm. This is why the whole first row is green, with the exception of the last cell.
%The only exception to this is if the method was unmatched in \cm. For this case, since we could not detect what type of change the method has undergone in \cm, we cannot reapply it to \an.
This leads us to discussing the last row and last column.
If a given method is unmatched in either changesets, then we conservatively assume that this is an integration that cannot be automated, and mark the cell as red.
The simplest example is if the method gets deleted in one changeset but not the other.
However, there may be more complicated types of changes that our tooling could not map.
Therefore, we choose the more conservative assumption that these changes cannot be safely automated.
Note that with the exception of the cell in the bottom right corer, all cells across the diagonal are marked as yellow, since if the same type of change is applied in both change sets, the feasibility of automation depends on the particular case.
For example, if both change sets include an argument or method rename, then the integration of the changes can be automated if both change sets used the same destination name. On the other hand, if the destination names are different, then developer intervention is needed to decide which name to choose.
We now discuss our reasoning for the color choice of the remaining cells, column by column, without repeating the reasoning for the first row, last row, last column, and diagonal in the table.

%\subsubsection{\ao $\rightarrow$ \cm Identical Methods}
%This row of the table is all green. The reason is that the changes that were applied to the \cm can all be applied to \an as well, since these mathods have not changed compared to \ao.

\textbf{\ao $\rightarrow$ \cm Method Move.}
%We can apply this kind of change in \cm to \an, if the method in \an is identical to \ao. This is true for any other type of change, hence the first row of the table is all green, except the last column which we will discuss later.
%If the method was not found is \an, it is not possible to apply this change. As this is true for any type of change, the bottom row of the table is colored red.
This column looks at cases where the method was moved in \cm.
%If the method was also moved in \an, then we cannot automatically determine which move should be applied, unless the method was coincidentally moved to the same place in both \an and \cm. Thus, this cell is marked as yellow.
If the method was inlined in \an, then it no longer exists and thus cannot be automatically moved, resulting in a red cell.
If part of the method was extracted in \an, then a tool cannot automatically decide whether it should create the extracted method in the same location as that in \an or if the extracted method should also be moved to the new location in \cm.
Because this requires the developer's decision, we color this cell red.
If the arguments or body of the corresponding method in \an are changed, the move change from \cm could still be easily applied in \an since the move operation does not interfere with any of these types of changes.

\textbf{\ao $\rightarrow$ \cm Method Rename.}
This change is very similar to Method Move. Thus, the coloring of most cells follows the same reasoning. The only exception is Method Extract in \an. If a method was extracted in \an and was renamed in \cm, a tool will not face the same question about where should it move the extracted part. The method will be extracted in the same location as \an and then renamed similar to \cm, thus this cell is colored green. When automating the integration of this change to \an, all calls to the method in \an should also be updated.

\textbf{\ao $\rightarrow$ \cm Method Inline.}
If a method is inlined in \cm, it means that its body was copied to all its call sites that existed in \ao, and the method itself has been removed. In this change, the implementation of the method is not altered, therefore, if there has been a body-only change in \an, we can still apply this change to the inlined code. However, any other type of change in \an that alters the method's signature cannot be easily automatically integrated. For example, assume that some argument got removed from the method's signature, applying this change to the inlined body of the method may not be that straightforward since it would require removing parts of the code in which the old method got inlined, an operation that is best left to the developer's judgment. The same logic applies to the methods that were inlined in \an, hence the (\ao $\rightarrow$ \an Method Inline) row is also mostly red.

\textbf{\ao $\rightarrow$ \cm Method Extract.}
This change occurs when part of the method's implementation in \ao is moved to a newly created method in \cm. If the method is moved in \an, a tool would not be able to automatically determine where to create the extracted method and the choice depends on the developer's decision. Since method extract affects a method's body, a tool would not be able to automatically apply the change in cases where \an also altered the body (those marked in red). All the other change categories do not alter a method's body, and thus, can be automatically integrated.~%TODO: \sn{but doesn't argument rename change the body too?}

\textbf{\ao $\rightarrow$ \cm Argument Rename.}
Argument renames can be automatically reapplied to \an in most cases. The only exceptions are when an argument is added or removed in \an, since the changed argument could potentially have the same name of the renamed argument in \cm. We mark these cases as yellow.

\textbf{\ao $\rightarrow$ \cm Argument Reorder.}
This change type is similar to the previous argument rename. The only difference is when a new argument is added or removed in \an. In this case, a tool cannot automatically determine how to reorder the arguments in the presence of a new argument or absence of a previously present argument. We mark both these cases as red.

\textbf{\ao $\rightarrow$ \cm Argument Add and Remove.}
We discuss both these columns together since their reasoning is similar.
When a new argument is added or an old one is deleted, a method's body is often altered as well to reflect the change. Therefore, if the method body is also modified in \an through any type of change, it is not possible to automatically reapply the changes from both change sets. If an argument is renamed in \an, it would be a problem if the same argument is removed in \cm. We will face a similar problem if the new argument name in \an is the same as the new argument added in \cm. Hence, we mark both these cases as yellow.

\textbf{\ao $\rightarrow$ \cm Argument Type Change.}
We assume that when an argument type change happens in \cm, the body is also changed.
Argument type changes can be automatically reapplied for changes in \an that only affect the method's signature, i.e. method rename, method move, and argument reorder. Although argument rename in \an mostly likely changes the method body, a tool can still automatically reapply the argument type change in \cm, because argument rename is merely a refactoring change than can be reapplied even if other parts of the body are changed.  
Other change types in \an involve non-trivial changes in the method's body that are likely to cause conflicts with argument type change in \cm. 

\textbf{\ao $\rightarrow$ \cm Body-only.}
The reasoning for this change type is very similar to the previous case, since we assume an argument type change also means a change in the method's body. The only difference is for the case where the method is inlined in \an. The body-only change from \cm can be re-applied by simply removing the old inlined version of the method and re-performing the inline operation with the new body from \cm.

\section{Evaluation Setup} \label{eval}

\begin{table}[t!]
\centering
\bgroup
\def\arraystretch{1.2}
\caption{Java subsystems in each \cs}
\label{table:dataset_stats}
\resizebox{0.4\textwidth}{!}{
\begin{tabular}{|c|c|c|c|c|} 
\hline
\csC & \ao & \an & \cm & \# of subsystems \\
\hline
CS1 & 4.2.2\_r1 & 4.3.1\_r1 & cm-10.1 & 208  \\ \hline
CS2 & 4.3.1\_r1 & 4.4.4\_r1 & cm-10.2 & 219  \\ \hline
CS3 & 4.4.4\_r2 & 5.0.2\_r1 & cm-11.0 & 275  \\ \hline
CS4 & 5.0.2\_r1 & 5.1.1\_r1 & cm-12.0 & 346  \\ \hline
CS5 & 5.1.1\_r37 & 6.0.1\_r1 & cm-12.1 & 355  \\ \hline
CS6 & 6.0.1\_r81 & 7.0.0\_r1 & cm-13.0 & 381  \\ \hline
CS7 & 7.0.0\_r14 & 7.1.2\_r1 & cm-14.0 & 392\\ \hline
CS8 & 7.1.2\_r36 & 8.0.0\_r1 & cm-14.1 & 398 \\ \hline

\end{tabular}
}
\egroup
\vspace{-0.5cm}
\end{table}

\paragraph{Identifying \cs{s}} We consider the \checkNum{eight} latest \lineage versions in our evaluation. To setup a \cs, we need the AOSP version each \lineage version is based on and the subsequent version of AOSP that it will need to be updated to.
Each version of \lineage is based on a corresponding version of AOSP, but it is not necessarily based on the first release of that AOSP version.
\lineage does not modify all of the repositories used in AOSP. 
Since these repositories are still required to build the source tree, \lineage fetches them from the AOSP servers.  
Fortunately, the exact AOSP version and release number for those unmodified repositories is mentioned in the manifest file for \lineage that contains the list of required repositories.
We assume that the repositories that \lineage modifies are also based on this documented release number, because it only makes sense for \lineage developers to include the unmodified AOSP repositories from the same version that modified repositories were based on; otherwise, incompatibilities may occur.

\paragraph{Downloading repositories \& identifying subsystems}
We are only interested in the mutual repositories between each version of \lineage and its corresponding AOSP version, since there is no point in studying repositories in \lineage that do not have a corresponding repository in AOSP and vice versa.
After processing the manifest file that contains the list of required repositories for each version of \lineage, our tooling automatically downloads the mutual repositories between \lineage and AOSP. It recursively searches all sub-folders for \textit{AndroidManifest.xml}, and considers such folders as subsystems. 
Table \ref{table:dataset_stats} shows the number of Java subsystems we analyzed for each \cs.

\paragraph{Performing the analysis} For each \cs, we apply our methodology in Section \ref{methodology}. 
For each subsystem in every \cs, we produce an output file that consists of the number of methods in each of the three versions of a given \cs (\ao, \an and \cm), as well as the number of methods in each overlapping category of changes, similar to Table \ref{table:methods_mappings}.

\section{Results} \label{results}
We now answer our three research questions.

\begin{table}[t!]
\centering
\bgroup
\def\arraystretch{1.2}
\caption{Most changed subsystems in \an and \cm}%for each \cs}
\label{table:most_changed_subsystems}
\resizebox{0.45\textwidth}{!}{
\begin{tabular}{|c|c|c|c|} 
\hline
Comparison & Number of changed & Number of changed  & Number of mutually \\
scenario & subsystems in \an & subsystems in \cm & changed subsystems  \\ \hline

CS1 &  51 &  23 &  20 (86.96\% of \cm) \\ \hline

CS2 &  49 & 26 & 22 (84.62\% of \cm) \\ \hline

CS3 &74 &  34 & 31 (91.18\% of \cm) \\ \hline

CS4 & 79 & 36 &  27 (75.0\% of \cm) \\ \hline

CS5 & 88 & 40 & 36 (90.0\% of \cm) \\ \hline

CS6 & 102 & 61 & 54 (88.52\% of \cm) \\ \hline

CS7 & 74 & 34 &23 (67.65\% of \cm) \\ \hline

CS8 & 84 & 33 & 26 (78.79\% of \cm) \\ \hline

\end{tabular}
}
\egroup
\end{table}

\subsection{Frequently Modified Subsystems} \label{results_rq1}

\begin{table*}
\centering
\bgroup
% \vspace{-0.5cm}
\def\arraystretch{1.2}
\caption{Heat map of average proportion of change types across all \cs{s}}
\label{plot:results_heatmap}
% \vspace{2mm}
\resizebox{0.67\textwidth}{!}{
\begin{tabular}{cc|c|c|c|c|c|c|c|c|c|c|c|c|c|} 
\hhline{~~|-|-|-|-|-|-|-|-|-|-|-}
 & & \multicolumn{11}{c|}{\textbf{\ao$\rightarrow$\cm} Changeset}\\ \hhline{~~|-|-|-|-|-|-|-|-|-|-|-}
& & \vcell{Method Move} & \vcell{Method Rename} & \vcell{Method Inline} & \vcell{Method Extract} & \vcell{Argument Rename} & \vcell{Argument Reorder} & \vcell{Argument Add} & \vcell{Argument Remove} & \vcell{Argument Type Change} & \vcell{Body-only} & \vcell{Unmatched} \\ \hline

\multicolumn{1}{|c|}{\parbox[t]{2mm}{\multirow{12}{*}{\rotatebox[origin=c]{90}{\textbf{\ao$\rightarrow$\an} Changeset}}}}
 & Identical & \cellcolor[HTML]{FEFEFE} \textcolor[HTML]{38AA5C}{0.28\%} & \cellcolor[HTML]{FEFEFE} \textcolor[HTML]{38AA5C}{0.4\%} & \cellcolor[HTML]{FEFEFE} \textcolor[HTML]{38AA5C}{0.06\%} & \cellcolor[HTML]{FCFCFC} \textcolor[HTML]{38AA5C}{1.25\%} & \cellcolor[HTML]{FEFEFE} \textcolor[HTML]{38AA5C}{0.11\%} & \cellcolor[HTML]{FEFEFE} \textcolor[HTML]{38AA5C}{0.1\%} & \cellcolor[HTML]{FBFBFB} \textcolor[HTML]{38AA5C}{2.09\%} & \cellcolor[HTML]{FEFEFE} \textcolor[HTML]{38AA5C}{0.31\%} & \cellcolor[HTML]{FEFEFE} \textcolor[HTML]{38AA5C}{0.35\%} & \cellcolor[HTML]{A9A9A9} \textcolor[HTML]{38AA5C}{50.19\%} & \cellcolor[HTML]{F1F1F1} \textcolor[HTML]{EA0D30}{8.21\%} \\ \hhline{|~|-|-|-|-|-|-|-|-|-|-|-|-}

\multicolumn{1}{|c|}{} & Method Move & \cellcolor[HTML]{FFFFFF} \textcolor[HTML]{FCC168}{0.0\%} & \cellcolor[HTML]{FEFEFE} \textcolor[HTML]{38AA5C}{0.01\%} & \cellcolor[HTML]{FFFFFF} \textcolor[HTML]{EA0D30}{0.0\%} & \cellcolor[HTML]{FFFFFF} \textcolor[HTML]{EA0D30}{0.0\%} & \cellcolor[HTML]{FFFFFF} \textcolor[HTML]{38AA5C}{0.0\%} & \cellcolor[HTML]{FFFFFF} \textcolor[HTML]{38AA5C}{0.0\%} & \cellcolor[HTML]{FFFFFF} \textcolor[HTML]{38AA5C}{0.0\%} & \cellcolor[HTML]{FFFFFF} \textcolor[HTML]{38AA5C}{0.0\%} & \cellcolor[HTML]{FFFFFF} \textcolor[HTML]{38AA5C}{0.0\%} & \cellcolor[HTML]{FEFEFE} \textcolor[HTML]{38AA5C}{0.05\%} & \cellcolor[HTML]{FEFEFE} \textcolor[HTML]{EA0D30}{0.0\%} \\ \hhline{|~|-|-|-|-|-|-|-|-|-|-|-|-}

\multicolumn{1}{|c|}{} & Method Rename & \cellcolor[HTML]{FFFFFF} \textcolor[HTML]{38AA5C}{0.0\%} & \cellcolor[HTML]{FEFEFE} \textcolor[HTML]{FCC168}{0.03\%} & \cellcolor[HTML]{FFFFFF} \textcolor[HTML]{EA0D30}{0.0\%} & \cellcolor[HTML]{FEFEFE} \textcolor[HTML]{38AA5C}{0.01\%} & \cellcolor[HTML]{FFFFFF} \textcolor[HTML]{38AA5C}{0.0\%} & \cellcolor[HTML]{FFFFFF} \textcolor[HTML]{38AA5C}{0.0\%} & \cellcolor[HTML]{FEFEFE} \textcolor[HTML]{38AA5C}{0.04\%} & \cellcolor[HTML]{FEFEFE} \textcolor[HTML]{38AA5C}{0.0\%} & \cellcolor[HTML]{FFFFFF} \textcolor[HTML]{38AA5C}{0.0\%} & \cellcolor[HTML]{FEFEFE} \textcolor[HTML]{38AA5C}{0.21\%} & \cellcolor[HTML]{FEFEFE} \textcolor[HTML]{EA0D30}{0.06\%} \\ \hhline{|~|-|-|-|-|-|-|-|-|-|-|-|-}

\multicolumn{1}{|c|}{} & Method Inline & \cellcolor[HTML]{FFFFFF} \textcolor[HTML]{EA0D30}{0.0\%} & \cellcolor[HTML]{FEFEFE} \textcolor[HTML]{EA0D30}{0.0\%} & \cellcolor[HTML]{FEFEFE} \textcolor[HTML]{FCC168}{0.01\%} & \cellcolor[HTML]{FFFFFF} \textcolor[HTML]{EA0D30}{0.0\%} & \cellcolor[HTML]{FFFFFF} \textcolor[HTML]{EA0D30}{0.0\%} & \cellcolor[HTML]{FFFFFF} \textcolor[HTML]{EA0D30}{0.0\%} & \cellcolor[HTML]{FFFFFF} \textcolor[HTML]{EA0D30}{0.0\%} & \cellcolor[HTML]{FFFFFF} \textcolor[HTML]{EA0D30}{0.0\%} & \cellcolor[HTML]{FFFFFF} \textcolor[HTML]{EA0D30}{0.0\%} & \cellcolor[HTML]{FEFEFE} \textcolor[HTML]{38AA5C}{0.06\%} & \cellcolor[HTML]{FEFEFE} \textcolor[HTML]{EA0D30}{0.01\%} \\ \hhline{|~|-|-|-|-|-|-|-|-|-|-|-|-}

\multicolumn{1}{|c|}{} & Method Extract & \cellcolor[HTML]{FEFEFE} \textcolor[HTML]{EA0D30}{0.0\%} & \cellcolor[HTML]{FEFEFE} \textcolor[HTML]{38AA5C}{0.0\%} & \cellcolor[HTML]{FFFFFF} \textcolor[HTML]{EA0D30}{0.0\%} & \cellcolor[HTML]{FEFEFE} \textcolor[HTML]{FCC168}{0.1\%} & \cellcolor[HTML]{FFFFFF} \textcolor[HTML]{38AA5C}{0.0\%} & \cellcolor[HTML]{FFFFFF} \textcolor[HTML]{38AA5C}{0.0\%} & \cellcolor[HTML]{FEFEFE} \textcolor[HTML]{EA0D30}{0.03\%} & \cellcolor[HTML]{FEFEFE} \textcolor[HTML]{EA0D30}{0.01\%} & \cellcolor[HTML]{FEFEFE} \textcolor[HTML]{EA0D30}{0.01\%} & \cellcolor[HTML]{FDFDFD} \textcolor[HTML]{EA0D30}{0.69\%} & \cellcolor[HTML]{FEFEFE} \textcolor[HTML]{EA0D30}{0.02\%} \\ \hhline{|~|-|-|-|-|-|-|-|-|-|-|-|-}

\multicolumn{1}{|c|}{} & Argument Rename & \cellcolor[HTML]{FFFFFF} \textcolor[HTML]{38AA5C}{0.0\%} & \cellcolor[HTML]{FFFFFF} \textcolor[HTML]{38AA5C}{0.0\%} & \cellcolor[HTML]{FFFFFF} \textcolor[HTML]{EA0D30}{0.0\%} & \cellcolor[HTML]{FEFEFE} \textcolor[HTML]{38AA5C}{0.01\%} & \cellcolor[HTML]{FEFEFE} \textcolor[HTML]{FCC168}{0.0\%} & \cellcolor[HTML]{FFFFFF} \textcolor[HTML]{38AA5C}{0.0\%} & \cellcolor[HTML]{FEFEFE} \textcolor[HTML]{FCC168}{0.02\%} & \cellcolor[HTML]{FFFFFF} \textcolor[HTML]{FCC168}{0.0\%} & \cellcolor[HTML]{FFFFFF} \textcolor[HTML]{38AA5C}{0.0\%} & \cellcolor[HTML]{FEFEFE} \textcolor[HTML]{38AA5C}{0.03\%} & \cellcolor[HTML]{FFFFFF} \textcolor[HTML]{EA0D30}{0.0\%} \\ \hhline{|~|-|-|-|-|-|-|-|-|-|-|-|-}

\multicolumn{1}{|c|}{} & Argument Reorder & \cellcolor[HTML]{FFFFFF} \textcolor[HTML]{38AA5C}{0.0\%} & \cellcolor[HTML]{FFFFFF} \textcolor[HTML]{38AA5C}{0.0\%} & \cellcolor[HTML]{FFFFFF} \textcolor[HTML]{EA0D30}{0.0\%} & \cellcolor[HTML]{FFFFFF} \textcolor[HTML]{38AA5C}{0.0\%} & \cellcolor[HTML]{FFFFFF} \textcolor[HTML]{38AA5C}{0.0\%} & \cellcolor[HTML]{FFFFFF} \textcolor[HTML]{FCC168}{0.0\%} & \cellcolor[HTML]{FEFEFE} \textcolor[HTML]{EA0D30}{0.01\%} & \cellcolor[HTML]{FEFEFE} \textcolor[HTML]{EA0D30}{0.0\%} & \cellcolor[HTML]{FFFFFF} \textcolor[HTML]{38AA5C}{0.0\%} & \cellcolor[HTML]{FEFEFE} \textcolor[HTML]{38AA5C}{0.02\%} & \cellcolor[HTML]{FFFFFF} \textcolor[HTML]{EA0D30}{0.0\%} \\ \hhline{|~|-|-|-|-|-|-|-|-|-|-|-|-}

\multicolumn{1}{|c|}{} & Argument Add & \cellcolor[HTML]{FEFEFE} \textcolor[HTML]{38AA5C}{0.04\%} & \cellcolor[HTML]{FFFFFF} \textcolor[HTML]{38AA5C}{0.0\%} & \cellcolor[HTML]{FFFFFF} \textcolor[HTML]{EA0D30}{0.0\%} & \cellcolor[HTML]{FEFEFE} \textcolor[HTML]{EA0D30}{0.03\%} & \cellcolor[HTML]{FEFEFE} \textcolor[HTML]{FCC168}{0.0\%} & \cellcolor[HTML]{FFFFFF} \textcolor[HTML]{EA0D30}{0.0\%} & \cellcolor[HTML]{FEFEFE} \textcolor[HTML]{FCC168}{0.18\%} & \cellcolor[HTML]{FEFEFE} \textcolor[HTML]{EA0D30}{0.01\%} & \cellcolor[HTML]{FFFFFF} \textcolor[HTML]{EA0D30}{0.0\%} & \cellcolor[HTML]{FEFEFE} \textcolor[HTML]{EA0D30}{0.32\%} & \cellcolor[HTML]{FEFEFE} \textcolor[HTML]{EA0D30}{0.1\%} \\ \hhline{|~|-|-|-|-|-|-|-|-|-|-|-|-}

\multicolumn{1}{|c|}{} & Argument Remove & \cellcolor[HTML]{FFFFFF} \textcolor[HTML]{38AA5C}{0.0\%} & \cellcolor[HTML]{FEFEFE} \textcolor[HTML]{38AA5C}{0.0\%} & \cellcolor[HTML]{FEFEFE} \textcolor[HTML]{EA0D30}{0.01\%} & \cellcolor[HTML]{FEFEFE} \textcolor[HTML]{EA0D30}{0.02\%} & \cellcolor[HTML]{FFFFFF} \textcolor[HTML]{FCC168}{0.0\%} & \cellcolor[HTML]{FEFEFE} \textcolor[HTML]{EA0D30}{0.01\%} & \cellcolor[HTML]{FEFEFE} \textcolor[HTML]{EA0D30}{0.07\%} & \cellcolor[HTML]{FEFEFE} \textcolor[HTML]{FCC168}{0.03\%} & \cellcolor[HTML]{FFFFFF} \textcolor[HTML]{EA0D30}{0.0\%} & \cellcolor[HTML]{FEFEFE} \textcolor[HTML]{EA0D30}{0.25\%} & \cellcolor[HTML]{FEFEFE} \textcolor[HTML]{EA0D30}{0.03\%} \\ \hhline{|~|-|-|-|-|-|-|-|-|-|-|-|-}

\multicolumn{1}{|c|}{} & Argument Type Change & \cellcolor[HTML]{FFFFFF} \textcolor[HTML]{38AA5C}{0.0\%} & \cellcolor[HTML]{FFFFFF} \textcolor[HTML]{38AA5C}{0.0\%} & \cellcolor[HTML]{FFFFFF} \textcolor[HTML]{EA0D30}{0.0\%} & \cellcolor[HTML]{FFFFFF} \textcolor[HTML]{EA0D30}{0.0\%} & \cellcolor[HTML]{FFFFFF} \textcolor[HTML]{38AA5C}{0.0\%} & \cellcolor[HTML]{FFFFFF} \textcolor[HTML]{38AA5C}{0.0\%} & \cellcolor[HTML]{FEFEFE} \textcolor[HTML]{EA0D30}{0.02\%} & \cellcolor[HTML]{FEFEFE} \textcolor[HTML]{EA0D30}{0.01\%} & \cellcolor[HTML]{FEFEFE} \textcolor[HTML]{FCC168}{0.02\%} & \cellcolor[HTML]{FEFEFE} \textcolor[HTML]{EA0D30}{0.1\%} & \cellcolor[HTML]{FEFEFE} \textcolor[HTML]{EA0D30}{0.04\%} \\ \hhline{|~|-|-|-|-|-|-|-|-|-|-|-|-}

\multicolumn{1}{|c|}{} & Body-only & \cellcolor[HTML]{FEFEFE} \textcolor[HTML]{38AA5C}{0.07\%} & \cellcolor[HTML]{FEFEFE} \textcolor[HTML]{38AA5C}{0.05\%} & \cellcolor[HTML]{FEFEFE} \textcolor[HTML]{38AA5C}{0.01\%} & \cellcolor[HTML]{FEFEFE} \textcolor[HTML]{EA0D30}{0.46\%} & \cellcolor[HTML]{FEFEFE} \textcolor[HTML]{38AA5C}{0.0\%} & \cellcolor[HTML]{FEFEFE} \textcolor[HTML]{38AA5C}{0.03\%} & \cellcolor[HTML]{FEFEFE} \textcolor[HTML]{EA0D30}{0.42\%} & \cellcolor[HTML]{FEFEFE} \textcolor[HTML]{EA0D30}{0.09\%} & \cellcolor[HTML]{FEFEFE} \textcolor[HTML]{EA0D30}{0.02\%} & \cellcolor[HTML]{E3E3E3} \textcolor[HTML]{FCC168}{16.09\%} & \cellcolor[HTML]{FDFDFD} \textcolor[HTML]{EA0D30}{0.7\%} \\ \hhline{|~|-|-|-|-|-|-|-|-|-|-|-|-}

\multicolumn{1}{|c|}{} & Unmatched & \cellcolor[HTML]{FEFEFE} \textcolor[HTML]{EA0D30}{0.02\%} & \cellcolor[HTML]{FEFEFE} \textcolor[HTML]{EA0D30}{0.07\%} & \cellcolor[HTML]{FEFEFE} \textcolor[HTML]{EA0D30}{0.01\%} & \cellcolor[HTML]{FEFEFE} \textcolor[HTML]{EA0D30}{0.15\%} & \cellcolor[HTML]{FEFEFE} \textcolor[HTML]{EA0D30}{0.02\%} & \cellcolor[HTML]{FEFEFE} \textcolor[HTML]{EA0D30}{0.01\%} & \cellcolor[HTML]{FEFEFE} \textcolor[HTML]{EA0D30}{0.46\%} & \cellcolor[HTML]{FEFEFE} \textcolor[HTML]{EA0D30}{0.12\%} & \cellcolor[HTML]{FEFEFE} \textcolor[HTML]{EA0D30}{0.02\%} & \cellcolor[HTML]{F0F0F0} \textcolor[HTML]{EA0D30}{8.43\%} & \cellcolor[HTML]{F3F3F3} \textcolor[HTML]{EA0D30}{6.67\%} \\ \hline

\end{tabular}
}
\egroup
\end{table*}

%\begin{center}
%\textit{RQ1: Which parts of Android are frequently modified in AOSP vs. \lineage?}
%\end{center}

In order to better understand the kind of changes AOSP or \lineage undergo, our first step, reflected in RQ1, is to look at the frequently modified subsystems.
% Understanding the nature of these subsystems, and if there are commonalities between AOSP and \lineage, helps us understand the reasoning behind changes. 
Table~\ref{table:most_changed_subsystems} shows the number of changed subsystems by \cm and \an in each comparison scenario. Column 2 shows the number of changed subsystems in \an, column 3 shows the number of changed subsystems in \cm, and column 4 shows the number of mutually changed subsystems between \an and \cm as well as the percentage with respect to \cm. 

Table~\ref{table:most_changed_subsystems} shows that the vast majority of subsystems changed by \cm are also changed by \an. On average across the comparison scenarios, approximately \checkNum{83\%} of the subsystems changed by \cm are also changed by \an. This high number illustrates the extent of the Android update problem.
However, the fact that a subsystem has been modified in \cm and \an does not tell us anything about the nature or extent of these changes. To better understand the changes that subsystems undergo, we extract the five most changed subsystems in \cm and \an for each \cs. We find that there are at least two mutually changed subsystems between \an and \cm in all \cs{s}. This suggests that AOSP and \lineage developers often apply a large number of changes to the same Android components. To better understand why developers need to change these subsystems, we further investigate two subsystems that seem to be commonly modified by both AOSP and \lineage developers, \textit{SystemUI} and \textit{Settings}.

% \begin{figure*}[t!]
% \centering
% \includegraphics[width=14.5cm]{plots/trends_plot.pdf}
% \caption{Percentage of \cm changes that have an overlapping change in \an across different \cs{s}}
% \label{plot:results_trends}
% \end{figure*}

\paragraph{SystemUI}
This subsystem is responsible for core user visible components that are generally accessible from any app, e.g. the status bar, volume control sliders, etc. The status bar can be expanded by dragging it down from the top of the screen. In the expanded mode, it provides useful features such as the list of notifications and quick settings. Since it is frequently used by Android users, it makes sense that AOSP developers frequently make changes to it. Many of \lineage-specific features are implemented in this subsystem, e.g., customizable quick settings. In Section \ref{results_rq2}, we discuss a method change sample from this subsystem which demonstrates how \lineage modifies the code to implement new features.

\paragraph{Settings}
The Settings subsystem is the interface that lets users reconfigure different settings in their Android device. There are several options available in Settings, often grouped into categories such as Connectivity or Display. AOSP developers often change this subsystem to amend new features and improve the user interface of existing items within each category. Settings in \lineage includes all options from AOSP, plus some additional \lineage-specific items that allow users to further customize their device. For example, there is an option under the \textit{Battery} category in \lineage for profiling CPU usage. Users can change this option to determine the workload on the CPU and subsequently its battery usage.

%\paragraph{Telephony}
%This subsystem is responsible for telecommunication functionality, such as making phone calls or sending and receiving text messages. It also provides APIs for third-party app developers to perform various actions regarding its functionality. Modifications by AOSP are mostly for implementing new protocols and bug fixes. \cm modifications are generally more specific, e.g., fixing a problem for a particular phone or service provider.

\begin{findingenv}{rq1}
On average, ~\checkNum{83\%} of the subsystems modified by \cm are also modified by \an. \textit{Settings} and \textit{SystemUI} are the two most commonly co-modified subsystems.
\end{findingenv}

\begin{figure*} %com.android.systemui.BatteryMeterView
\centering
\begin{minipage}[t]{.33\linewidth}
\begin{lstlisting}[title=\cm]
public void onDetachedFromWindow() {
    super.onDetachedFromWindow();
 <<+  mAttached = false;
    getContext().unregisterReceiver(mTracker);
 @@-  mBatteryController.removeStateChangedCallback(this);
 @@-  getContext().getContentResolver().
 @@-  		unregisterContentObserver(mSettingObserver);
 <<+  if (mBatteryStateRegistar != null) {
 <<+      mBatteryStateRegistar.removeStateChangedCallback(this);
 <<+  }
}
\end{lstlisting}
\end{minipage}
\begin{minipage}[t]{.33\linewidth}
\begin{lstlisting}[title=\ao]
public void onDetachedFromWindow() {
    super.onDetachedFromWindow();
    getContext().unregisterReceiver(mTracker);
    mBatteryController.removeStateChangedCallback(this);
    getContext().getContentResolver().
    		unregisterContentObserver(mSettingObserver);
}
\end{lstlisting}
\end{minipage}
\begin{minipage}[t]{.33\linewidth}
\begin{lstlisting}[title=\an]
public void onDetachedFromWindow() {
    super.onDetachedFromWindow();
 @@-  getContext().unregisterReceiver(mTracker);
    mBatteryController.removeStateChangedCallback(this);
 @@-  getContext().getContentResolver().
 @@-   		unregisterContentObserver(mSettingObserver);
 <<+  mDrawable.stopListening();
 <<+  TunerService.get(getContext()).removeTunable(this);
}
\end{lstlisting}
\end{minipage}
\vspace{-0.4cm}
\caption{Example method in \ao from \textit{SystemUI} subsystem that went through different body-only changes in \an and \cm. %This is an example of changes that cannot be automatically merged.
}
\label{figure:body-body-snippet}
\end{figure*}

\subsection{Overlapping Changes} \label{results_rq2}

%The mere existence of overlaps between subsystems that AOSP and \lineage modify in each version does not mean that automation for merging \cm changes to \an in such subsystems is impossible. 
In RQ2, we further study the types of changes that happen in mutually changed subsystems to better understand the nature of changes that are applied by AOSP developers, versus \lineage developers.
%We want to understand how AOSP and \lineage developers change the code, and whether there is a consistent trend across all \cs{s} for these types of changes.
To analyze this, we calculate the number of method changes and their types among all subsystems for each \cs. Table \ref{plot:results_heatmap} shows the average proportion of each change type across all \cs{s}.
The rows and columns, as well as the text colors, correspond to those in Table~\ref{table:methods_mappings}, making each cell correspond to an overlap cell in Table~\ref{table:methods_mappings}. The number in each cell indicates the percentage of the corresponding change overlap with respect to the total number of changes in \cm, as an average of all \cs{s}. Let us take the \ao$\rightarrow$\cm:Body-only and \ao$\rightarrow$\an:Body-only cell as an example. The number in the cell indicates that on average, \checkNum{16.09\%} of all methods that are changed in \cm had body-only changes in both \cm and \an.
For easier visualization, the table is drawn as a heat map, where the grayscale intensity of the cell background corresponds to the percentage value in the cell.

% CM:Body AN:Identical
The heat map suggests that the majority of the methods that are changed in \cm have body-only changes (Body-only column). Out of these methods, most of them remain identical in \an. Specifically, \checkNum{50.19\%} of the methods that were changed in \cm have body-only changes in \cm but remain identical in \an. %In absolute numbers, this corresponds to \checkNum{9,156} methods.
This is good news for automatically integrating \cm changes into \an, because there are no overlapping changes in \an. Combining this information with the information in Table~\ref{table:most_changed_subsystems}, it seems that while there is a very high overlap in terms of changed subsystems, there is less change overlap at the method level.

Given that on average almost half the methods changed in \cm (\checkNum{50.19\%}) remain identical in \an, we further investigate this cell from Table~\ref{plot:results_heatmap}. We search for the subsystem with the most number of changes in this particular category, across all \cs{s}. The \textit{packages\_apps\_Bluetooth} subsystem in CS7 matches this criteria. Out of \checkNum{602} \cm changes in this subsystem, \checkNum{508} of them remained identical in \ao and had body-only changes in \cm. We randomly sample \checkNum{20} of these \checkNum{508} methods and manually inspect them. We realize that for \checkNum{17} of these methods, there is either a new line added for logging an activity, or an existing line of code regarding a log message was edited. This suggests that the relatively higher number of body-only changes to methods in \cm in CS7 may be because of a new decision towards more thorough logging.

% CM:Body AN:Body changes
As Table~\ref{plot:results_heatmap} shows, the next highest category of changes is the body-only changes in both \cm and \an. 
%Such methods could have potential merge conflicts.
To better understand this category, we again search for the subsystem with the most changes in this category. It turns out to also be \textit{packages\_apps\_Bluetooth}, but this time in CS8 with \checkNum{170} such changes. Upon further inspection, we realize that similar to the previous case, most of these changes consist of adding a line for logging an activity in \cm, while a different body change has happened in \ao. 

In order to find a more interesting example, we picked the subsystem with the second most number of changes for this category. We find \textit{SystemUI} for CS6 with \checkNum{163} such changes. We look for changes that reflect different body-only change overlaps. Figure \ref{figure:body-body-snippet} shows code snippets from a sample method in \ao that has gone through body-only changes in both \an and \cm. 
The code snippet provides an interesting insight on what kind of changes \lineage and AOSP make. This method belongs to the \textit{BatteryMeterView.java} class which is responsible for displaying the battery status in the status bar. The method is called when the instance of the class is destroyed. In \ao, first the super method is called. Then, a receiver is unregistered. In the next line, the class is removed from the list of callbacks in \textit{BatteryController}. Finally, an observer is unregistered. In \an, the statements that unregister the receiver and the observer are deleted. However, calls to the super method and \textit{BatteryController} remain unchanged and two new lines are added. Upon further inspection, we find out that the new lines are related to new features in the new version of AOSP. In \cm, similar to \an, the call for unregistering the observer was removed. However, unlike \an, the call to \textit{BatteryController} was also removed and the statement for unregistering the receiver was not removed. A new statement was added in \cm that removes the class from the list of callbacks in \textit{BatteryStateRegistar} class. After exploring the code, we find out that \textit{BatteryStateRegistar} is \lineage's replacement for AOSP's \textit{BatteryController} and adds new options for customizing the battery icon style, something that is not implemented in AOSP.

\begin{findingenv}{rq2}
On average, \checkNum{~50\%} of changed \cm methods have body-only changes in \cm and remain identical in \an. Methods with body-only changes in both \cm and \an make up an average of \checkNum{~16\%} of changes in each \cs. %In most cases, the percentage of the overlapping changes follow the same trend across all \cs{s}.
\end{findingenv}

\subsection{Overall Feasibility of Automation} \label{results_rq3}

\begin{figure}[t!]
\centering
\includegraphics[width=8cm]{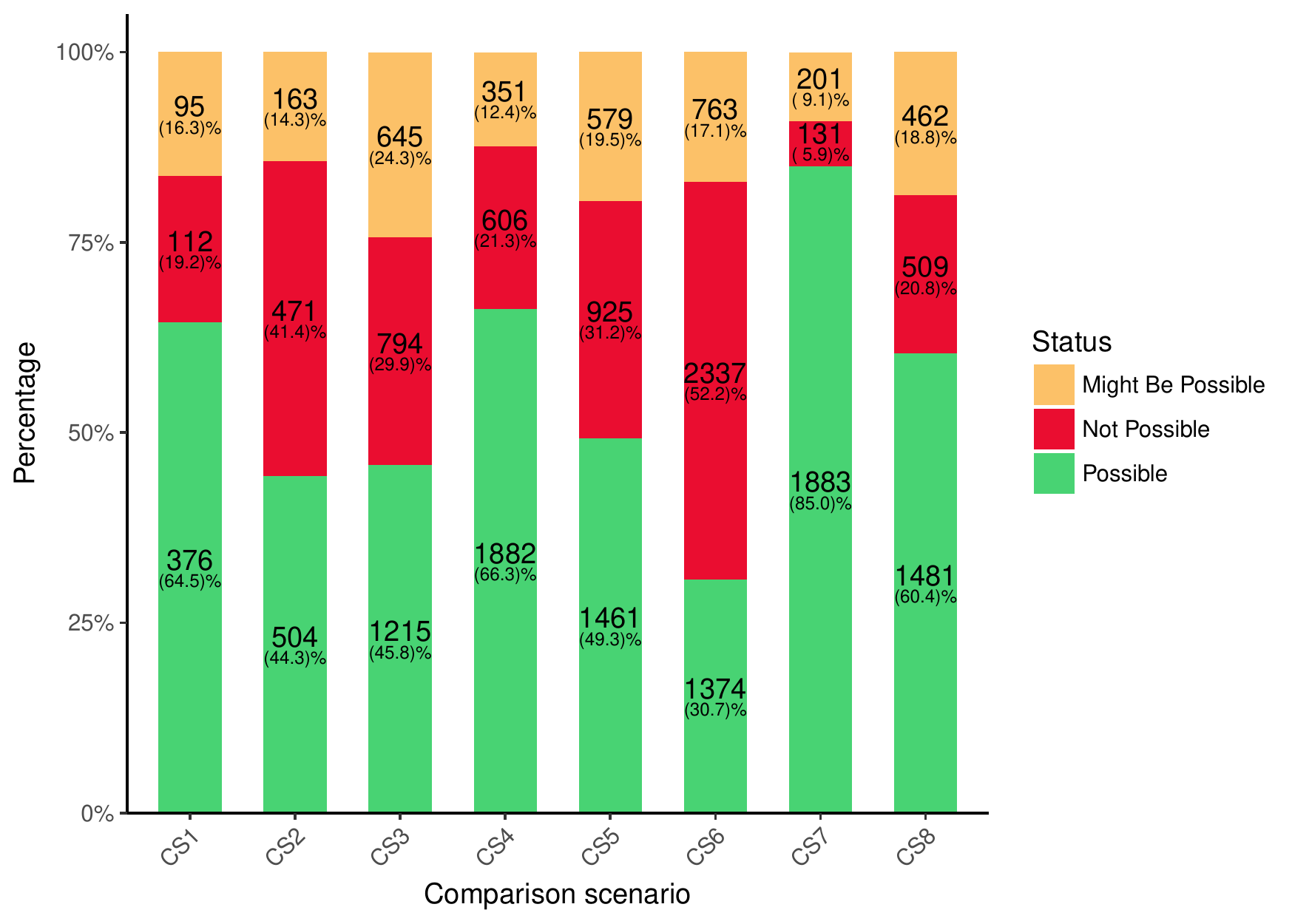}
\caption{Number and Percentage of \cm changed methods, categorized by feasibility of automated merging into \an for each \cs.\vspace{-0.5cm}}
\label{plot:results_bar_plot}
\end{figure}

RQ3 looks at the more general scale, where we are interested to know the percentage of all \cm changes that have the potential to be automatically integrated into \an. Based on Table~\ref{table:methods_mappings} and our reasoning behind each cell explained in Section \ref{sec:mapping-analysis}, we divide the aggregated analysis results for each \cs into three categories based on the potential for automation: \catGreen (Green), \catYellow (Yellow), \catRed (Red). Figure \ref{plot:results_bar_plot} visualizes the aggregated results of the analysis for each \cs. As the plot suggests, the majority of changes in \cm can automatically be re-applied onto \an for most of \cs{s}. On average, out of all \cm changes across each \cs, \checkNum{56\% (median: 55\%)} of them belong to the \catGreen category, \checkNum{28\% (median: 26\%)} belong to the \catRed category and the remaining \checkNum{16\% (median: 17\%)} belong to the \catYellow category.
%The plot also suggests that the same trend holds for each category of \cm changes across the 7 \cs{s} that we have. It is worth mentioning that we are conservatively assuming that all \ao methods that are deleted in both \an and \cm are duplicates and are thus purged. Because we believe that genuine changes in this category would pose no conflict to the automation, we expect the actual number for the percentage of changes in the No conflict category to be higher than the number reported here.

\begin{findingenv}{rq3}
On average, \checkNum{56\%} of \cm changes have the potential to be automatically merged into \an. An average of \checkNum{28\%} would need developer intervention, while an average of \checkNum{16\%} could \textit{potentially} be problematic.
\end{findingenv}

\section{Discussion} \label{sec:discussion}

This paper is a first step towards investigating the complexity of the Android update problem, and the potential for providing automated support for it. While we were mainly interested in the Android OS, given its size and prevalence, the methodology we used for investigating the problem can be used in other contexts. Our tool chain can be used to investigate the changes that happened in any Java software family where various variants of the software may exist and updates in one direction often need to be made. 

In specific to the Android update problem, our empirical investigation showed that while many subsystems are simultaneously modified in \lineage and AOSP, as many as \checkNum{63\%}, on average, of the methods changed in \lineage are not changed in AOSP. This is good news for solving the Android update problem. If we generally look at all the non-problematic categories in Table~\ref{table:methods_mappings}, an average of \checkNum{56\%} of method-level changes in \lineage can potentially be automatically integrated with AOSP changes with every new release. Our analysis of potential for automation was based on numerous discussions by the authors of this paper on whether a tool can handle the merge or integration without the developer's input. We relied on the semantics of the change to decide on the categorization of the overlap. This is because, as noted in the introduction, there currently does not exist a single off-the-shelf tool that can correctly and efficiently handle all the types of changes we discussed in this paper. In order to advance the state of the art and to develop tools and methods that can handle real-world integration and merge problems, such as those illustrated by the Android update problem studied, we dedicate the rest of this section to discuss how we envision using or extending current tooling or techniques to help with the Android update problem, as well as similar problems in other domains or contexts.

\subsection{`\catGreen' Categories}
\label{sec:no-conflict-disc}
In this section, we discuss how a practical tool could merge the categories of change in Table~\ref{table:methods_mappings} that have the potential to be automatically merged. The techniques discussed here resemble a three way merge, with \ao being the base and \an and \cm being the other two branches. The idea is that a tool should apply both sets of changes, from \an and \cm, to \ao. We only discuss the table cells with unique combinations of \an and \cm change types. For example, if we discuss the (\ao $\rightarrow$ \an Method Move) and (\ao $\rightarrow$ \cm Method Rename) cell, we do not discuss the (\ao $\rightarrow$ \an Method Rename) and (\ao $\rightarrow$ \cm Method Move) cell.

\textbf{\ao $\rightarrow$ \an Identical.}
\label{sec:no-conflict-identical}
We first consider the methods that were refactored in \cm (i.e. the method was moved, renamed, inlined, extracted, or its arguments were renamed or reordered). Since the method remained identical in \an, a simple textual merge, e.g., by git, would be enough to apply the change to that specific method. However, a more thorough way of merging the change is to apply the actual refactoring to make sure that references to the method can be updated accordingly. Since the change detection tools we use, i.e., RefactoringMiner and ChangeDistiller, can already identify the exact type of refactoring that occurred, refactoring parameters can be provided to existing refactoring engines, e.g., in Eclipse~\cite{eclipse} or IntelliJ IDEA~\cite{intelliJ}, to automatically apply the change.

Next, we discuss those methods that have an argument change in \cm (i.e. a new argument was added, and old argument was removed, or the type of an argument was changed). Using textual merge tools, the entire method in \an could be replaced with its corresponding method in \cm. At this point, the change is merged with no textual conflicts. However, to avoid compilation errors, references to the method also need to be updated. This can be done using static analysis, where all calls to the method could be found and updated to match the new method signature. In the case of argument deletion, updating the references is straightforward. In the case of argument addition, a candidate parameter can be added to the method call by statically finding a variable with the right type in that context. This is obviously a heuristic that may not always work correctly. Another option is to use executable transformations that are extracted from change distilling, similar to the technique suggested by Stevens and De Roover~\cite{StevensTransformations}. This way, we can extract an exact transformation script from the change detected in \cm, and execute the exact transformation on \ao.

Finally, we consider the methods that have changes only in their bodies in \cm. Textual merge tools can simply be used to execute the merge. Since there are no changes in the method's signature, there is no need to find and update the calls to the method. However, there may be a risk of test failures due to the updated functionality.

\textbf{\ao $\rightarrow$ \an Method Move.}
Using our change detection tools, the refactoring parameters for the method move in \an can be extracted. These parameters can be passed to a refactoring tool to reapply the method move to \ao.
Now, if the same method underwent an argument rename or argument reorder change in \cm, the extracted refactoring parameters from the \cm change can be updated to reflect the new location of the method. A refactoring engine can be then used to reapply the \cm change to the method in its new location in \ao. If the change in \cm was adding or removing an argument, changing an argument's type, or a body-only change, textual merge tools can be used to replace the entire method in its new location with the new implementation from \cm. Finally, the tool can utilize static analysis to update all references to the updated method.

\textbf{\ao $\rightarrow$ \an Method Rename.}
Similar to the previous category, the tool can extract the refactoring parameters for the rename in \an. Using a refactoring engine, this change can be applied to the original method $m1$ in \ao. Now, if part of $m1$ was extracted to $m2$ in \cm, textual merge can be used to integrate $m2$ into \ao. Next, the modified body of $m1$ in \cm can be replaced with its implementation in \ao. If $m1$ had a different change type in \cm, it can be merged to \ao similar to what we discussed in the previous category.

\textbf{\ao $\rightarrow$ \an Method Inline.}
Suppose that \an inlined the body of $m1$ into $m2$.
In this case, only body-only changes to $m1$ in \cm can be automatically integrated.
Textual merge can be used to integrate the \cm body-only changes of $m1$ into \ao.
The refactoring parameters from the \an inline change can then be extracted. These parameters include the signature and position of $m2$ that $m1$ was inlined into.  Using the extracted paramters, a refactoring engine can then be used to inline the updated $m1$ into $m2$ in \ao.

\textbf{\ao $\rightarrow$ \an Method Extract.}
Suppose that part of $m1$ was extracted into $m2$ in \an.
Two corresponding refactoring changes in \cm can be automatically integrated: argument rename or argument reorder. A tool could first apply the refactoring change from \cm to $m1$ in \ao. The refactoring parameters for the method extract change from \an can then be provided to a refactoring engine to extract the now updated body of $m1$ in \ao into $m2$.

\textbf{\ao $\rightarrow$ \an Argument Rename.}
This change type could be merged if the \cm change type is either argument reorder, argument type change, or body-only change. The tool can first merge the \cm change into \ao using textual merge tools. If there were any changes to the arguments in \cm, the refactoring parameters related to the argument rename change in \an should be updated to reflect the change from \cm. Next, the argument rename change can be applied using a refactoring engine and the new refactoring parameters.

\textbf{\ao $\rightarrow$ \an Argument Reorder.}
If the method's change type in \cm is body-only, the new implementation can be replaced in \ao using textual merge tools. Then, the method's signature in \an, that includes the new ordering of arguments, could be replaced with the signature in \ao. If the method's change type in \cm is argument type change, it can be applied to \ao using textual merge tools. Next, the refactoring parameters related to the argument reorder change in \an should be updated to reflect the new argument type \cm. Using a refactoring engine and the new refactoring parameters, the arguments can be reordered.

\subsection{`\catRed' Categories}
\label{sec:conflict-disc}
Unlike the previous category, the merging process of the methods in this category cannot be fully automated. This is mainly because the integration of the two changes in \an and \cm can only be completed based on the developer's decision. For example, if a method is deleted in \an and it is moved in \cm, it is hard to automatically decide if the method should be added back to \an or if the method move should be ignored since the method is deleted. A completely different resolution may also be decided by the developer. For example, she may decide to move the functionality in the refactored \cm method to a different method in \an. One option would be to at least create some analysis that would calculate potential options and show a preview of these options to the developer. Once the developer decides, the change resolution can potentially be automatically applied.

\subsection{`\catYellow' Categories}
For these categories, automation depends on the details of the change.
For example, for a method in \ao that had one of its arguments renamed in \an and a new argument was added to it in \cm, automation might not be possible. More specifically, if the name of the new argument in \an is the same as the new name of the renamed argument in \cm, automation is not possible.
However, if this is not the case, a tool may be able to automatically merge these changes.
Another case consists of methods in \ao that have undergone the same type of change in both \an and \cm. Integration of such changes can only be automated if the changes are identical. For example, if a method in \ao was moved in both \an and \cm, we can consider it as a change with possible automation only if it was moved to the same location in both \an and \cm.
%In some cases, such as a method rename in \an and an argument rename in \cm, automation may be possible by using a refactoring engine to perform the method rename and then applying the techniques described in the \textit{AN:identical CM:argument} category in Section~\ref{sec:no-conflict-disc}.
%However, if both refactoring changes move the method to different classes, a developer needs to decide how to merge the change in \cm to \an. She can either choose to apply the move made in \cm to \an or to leave the method in \an as it is. 
A practical solution to changes in this category is to create a list of all change combinations and determine the conditions under which they can be merged automatically.%, or require a decision by a developer, similar to the changes described in Section~\ref{sec:conflict-disc}. 

% \subsection{Prioritizing \lineage over AOSP}
% Note that our discussion of above tooling and the categories we created in Table \ref{table:methods_mappings} are based on 
% the premise that \cm changes have no priority over \an changes. One could argue that in case of conflicts, phone vendors would rather keep their proprietary changes over changes that have happened in AOSP. Thus, it would be interesting to investigate if automated tooling can be configured to prioritize one set of changes over the other. 
%We also investigated this case by prioritizing changes made by \lineage developers and reassigning colors to Table \ref{table:methods_mappings}. Under this condition, we observe an increase of~\checkNum{0.05\%} in the number of methods with possible automation. This suggests that prioritizing vendor-specific changes over \an changes will not yield any significant advantages. 

\section {Threats to Validity} \label{sec:threats}
% use of CynogenMod and not a real vendor
We used \lineage, an open-source community-driven Operating System based on AOSP, in our study. While we cannot generalize our findings to all other phone vendors, our methodology can be applied to other variants of Android, such as proprietary code of phone vendors. Moreover, \lineage has a large number of active users (50 million in 2015) and a community of more than 1,000 developers. We believe that these qualities render \lineage as a suitable alternative to a mainstream phone vendor for the purposes of our study. Additionally, the findings of our study already illustrate the problems that arise when independently modified versions of the same software system need to be merged. When searching for subjects, we did consider other variants of Android that are actively maintained, but ruled them out for different reasons. \tool{AOKP} has a considerable number of users, but is based on \lineage. \tool{MIUI} \cite{MIUI}'s source code was not entirely publicly available. While \tool{Paranoid Android} \cite{Paranoid} is fully open-source, we found that it does not have as many releases as \lineage. Also, the number of changes in each release were considerably lower than those in \lineage.

% method-level choice, test suites and configuration files
We only cover changes on the method-level. Although there are other types of changes such as changes to classes and their attributes, we believe that methods provide a good balance between fine-grained changes and at the same time a meaningful evaluation unit. Evolution studies at the method level have also been performed in several previous work~\cite{SMR:SMR344}.

% can we be missing any method-level changes
The correctness of the method-level changes we identify relies on the recall and precision of the tools we use: SourcererCC, RefactoringMiner, and ChangeDistiller. The tools we use are well established and have been previously used in several studies~\cite{changedistiller:usage, refactoringminer:usage}. Additionally, all tools have high reported recall and precision rates, which we provided in Section~\ref{methodology}.
We manually sampled several of their findings to verify that the tools work as expected.
%The only issue we faced was with SourcererCC, where we found that it misses some Java files during its indexing, and so does not report clones in these files which would lead to inaccuracies in identifying identical changes. However, we could not identify the reason for this.  We implemented a workaround for this problem by performing a string comparison between each pair of candidate methods for the body-only change type. If the comparison reveals that the methods are identical, we add them to the identical change type (instead of incorrectly considering them as body-only change).

% our categorization of conflicts is based on the semantics of each change type, but we did not actually perform the merge to check
Finally, we do not actually perform any merging or integration, and accordingly, we do not analyze if our anticipated automation may lead to build or test errors.
% This was beyond the scope of our current work since we focused on empirically identifying and understanding the types of changes that occur, and their overlap.
Given the size of Android, any merge attempt using current tools would most likely result in a considerable number of conflicts and resolving them is a daunting task. This is why we decided to first study the nature of changes that occur.
Our results show what kinds of change overlap occur in practice, and can be used to guide efforts in improving automation tools.
As our next step in our future work, we plan to investigate existing textual merge tools such as git, existing structural merge tools~\cite{Apel:2011SS,Apel:2012:SMA}, as well as the combination of tools mentioned in Section~\ref{sec:discussion} to practically apply the integration and analyze the results~\cite{eclipse,intelliJ,StevensTransformations,RefactoringMiner}.
\section{Conclusion} \label{conclusion}

We performed an empirical study to understand the details of the Android update problem. The problem occurs when a new version of the Android OS is released, and phone vendors need to figure out how to re-apply their proprietary modifications to the new version. This leads to late adoption of the new version of Android by phone vendors, which may lead to problems associated with using an outdated version of the OS, such as security vulnerabilities.

By studying \checkNum{eight} versions of a community-based variant of Android called \lineage, which we use as a proxy for a phone vendor, we analyzed the details and overlap of the changes applied in \lineage versus those in Android. Based on the semantics of the changes, we categorized whether the overlapping changes have the potential to be automatically integrated or not.

Our results show that both \lineage and Android often change similar parts of the OS. We find that the subsystems related to the settings in Android as well as the user interface are mutually changed by \lineage and Android. On the other hand, when taking the semantics of the changes into account, we find that the majority of overlapping changes (\checkNum{56\%}) have the potential to be automatically merged. We discussed concrete automation opportunities for all categories of change overlaps. 
Our results are a useful first step for solving the Android update problem, and also for solving the general problem of integrating independently modified variants of a software system. 
The tooling we used is open-source and can be applied to investigate overlapping changes in any Java system. %Our next step is to investigate the results of applying the existing tools we identified  suggested automated merging techniques and develop an integrated tool that can be used to automate this merging task.

% As the results in table \ref{table:results} suggest, the proportion of changes that can be automated does not follow the same trend in all subsystems. In order to get a better idea, we need to choose a larger sample of subsystems, or do it for the whole Android platform. This is feasible because our method does not involve any manual steps and is scalable.

% Another important conclusion is that the migration task cannot be automated entirely. Although 81.58\% of methods in our sample subsystems can be automatically migrated, there are a considerable number of methods that require manual migration.

% We are going to further investigate the situation in yellow cells in tables \ref{table:methods_settings} to \ref{table:methods_dialer}. These are the methods in A6 that have a different change in their body implementation in A7 and CM13. In order to explore the feasibility of automation for such changes, we are going to use source code difference extractor such as ChangeDistiller \cite{changedistiller}. Another approach is tracking the history of commits in the version control software for those methods in A7. We can generate a patch file from the changes made to those methods from A6. Applying the patch to CM13 will certainly cause conflicts, since the method's body is changed in CM13 as well. However, we can investigate techniques to mitigate these conflicts.

\bibliographystyle{ACM-Reference-Format}
\bibliography{bibliography} 

\end{document}